\documentclass[a4paper,11pt]{article}
\usepackage{pos}
\usepackage{multirow}

\title{Long-term monitoring of the radio-galaxy M87 in gamma-rays: joint analysis of MAGIC, VERITAS and Fermi-LAT data}
 \ShortTitle{Long-term monitoring of the radio-galaxy M87}

\author*[a]{M. Molero}
\author[b]{P. Batista}
\author[c]{L. Fortson}
\author[a]{M. Nievas Rosillo}
\author[b]{E. Pueschel}
\author[c]{D. Ribeiro}
\author[a]{M. V\'{a}zquez Acosta}

\affiliation[a]{Instituto de Astrof\'{i}sica de Canarias (IAC) and Dpto. de Astrof\'{i}sica, Universidad de La Laguna, E-38200, La Laguna, Tenerife, Spain}

\affiliation[b]{Deutsches Elektronen-Synchrotron (DESY), Platanenallee 6, 15738 Zeuthen, Germany}

\affiliation[c]{School of Physics and Astronomy, University of Minnesota, Minneapolis, MN 55455, USA}

\onbehalf{on behalf of the MAGIC and VERITAS Collaborations} 

\emailAdd{mmolerog@iac.es}

\abstract{M87 was discovered in the very-high-energy band (VHE, E > 100 GeV) with HEGRA in 2003, long before its emission was detected in the high-energy band (HE, E > 100 MeV) with Fermi-LAT in 2009, opening the window to a new family of extragalactic sources with tilted jets. After a series of major VHE flares in 2005, 2008, and 2010, which were detected in multiple bands, the source has been found in a low activity state, interrupted only by comparatively smaller-scale flares. MAGIC and VERITAS, two stereoscopic Cherenkov telescope arrays located at Roque de los Muchachos Observatory (Canary Islands, Spain) and the Fred Lawrence Whipple Observatory (Arizona, US), have monitored M87 continuously and in coordination for more than 10 years. In this work, we present the data for 4 years of MAGIC and VERITAS observations corresponding to 2019, 2020, 2021 and 2022. The resulting light curves are shown in daily and monthly scales where no significant variability is observed. In addition, we show the first joint analysis using combined event data from the two VHE instruments and Fermi-LAT to compute the spectral energy distribution.}

\ConferenceLogo{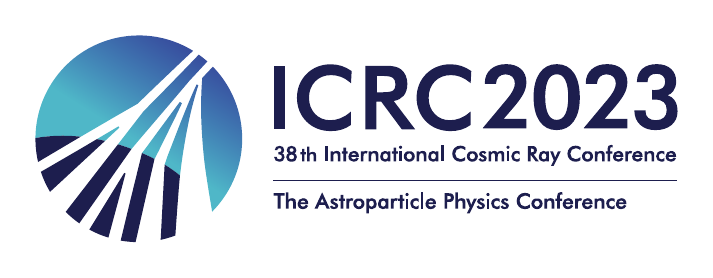}

\FullConference{%
38th International Cosmic Ray Conference (ICRC2023)\\
  26 July - 3 August, 2023\\
  Nagoya, Japan}

\begin{document}
\maketitle

\section{Introduction}

M87, also known as Messier 87, is a giant elliptical radio galaxy located in the Virgo Cluster, approximately 16 Mpc away \cite{Ref:M87Dist} and, it is classified as a Fanaroff-Riley-I-type (FR I) galaxy \cite{Ref:FRI}. The galaxy is powered by a supermassive black hole with a mass of (6.5 $\pm$ 0.7) $\times 10^9$ $M_{\odot}$  \cite{Ref:MassM87}. The jet, which extends to $\geq 30''$ \cite{Ref:JetM87Exten}, is misaligned with an angle between 15 $^\circ$ to 25 $^\circ$ \cite{Ref:JetM871,Ref:JetM872} with respect to our line of sight  and has been resolved in the radio, optical and X-ray bands \cite{Ref:JetM87Resol1,Ref:JetM87Resol2}.

M87 was first discovered in the very-high-energy band (VHE, E > 100 GeV) by the HEGRA collaboration in 2003 \cite{Ref:M87Hegra} and was later confirmed as a VHE emitter by H.E.S.S, MAGIC and VERITAS \cite{Ref:M87HESS,Ref:M87MAGIC,Ref:M87VTS}. In 2009, Fermi-LAT also detected M87 in the high-energy band (HE, E > 100 MeV) \cite{Ref:M87Fermi}. These observations established M87 as a HE and VHE gamma-ray emitter and opened the window to a new family of active galactic nuclei (AGN) with misaligned jets.

As of today, the source has been found in three different episodes of flare activity in the VHE band, in 2005, 2008 and 2010 \cite{Ref:M87HESS,Ref:M87MAGIC,Ref:M87VTS,Ref:M87All}, and in a quiescent low flux state since 2010 \cite{Ref:M87MAGIC2, Ref:M87EHT2017} with the exception of smaller-scale flares. 

In this context, MAGIC and VERITAS telescopes have periodically and in coordination monitored M87 since the last high activity period for more than 10 years. This data provides a unique opportunity to understand the evolution of the VHE emission of M87 during a low flux state.

\section{The MAGIC and VERITAS Telescopes}

MAGIC and VERITAS are two stereoscopic Imaging Atmospheric Cherenkov Telescopes (IACTs) arrays located at the Roque de los Muchachos Observatory (Canary Islands, Spain) and the Fred Lawrence Whipple Observatory (Arizona, US) respectively. The telescopes have been designed to detect the VHE gamma-ray emission from different galactic and extragalactic sources. The detection technique relies on the measurement of the Cherenkov light produced by the interaction of the gamma rays with the atmosphere. The main characteristics of both instruments are summarized in Table \ref{Tb:Telescopes}. 

\begin{table}[h]
\centering
\resizebox{\textwidth}{!}{%
\begin{tabular}{|c|c|c|c|c|c|c|}
\hline

\multirow{2}{*}{\textbf{Telescope}} & \textbf{Number of}  & \textbf{Diameter} & \textbf{Camera FoV} & \textbf{Pixel FoV} & \textbf{Number of}
& \textbf{Energy Range} \\

 & \textbf{Telescopes} & \textbf{{[}m{]}}  & \textbf{{[}$\boldsymbol{\deg}${]}} & \textbf{{[}$\boldsymbol{\deg}${]}} & \textbf{Pixels} & \textbf{{[}GeV{]}} \\ 

\hline
MAGIC      &2          & 17                        & 3.50                              & 0.10                             & 1039               & > 50                         \\ 
\hline
VERITAS      &4        & 12                        & 3.50                              &               0.15                  & 499                & > 85                         \\ 
\hline
\end{tabular}%
}
\caption{Summary of the main MAGIC and VERITAS telescope parameters \cite{Ref:MAGICPerformance, Ref:VTSPerformance}.}
\label{Tb:Telescopes}
\end{table}

\section{MAGIC and VERITAS Datasets and Analysis Method}

The M87 observations included in this work were performed from December to July of 2019, 2020, 2021 and 2022 respectively. The data were collected at zenith angles ranging from 15 $^\circ$ to 50 $^\circ$ and during dark and moonlight conditions. Then, only the data collected under good weather conditions were kept in the analysis. After the data quality cuts a total of 112 and 61 hours of effective observation time was obtained for MAGIC and VERITAS respectively. In addition, a summary of the effective observation time for each period and for both instruments is shown in Table \ref{Tb:Datasets}.

The MAGIC data were analyzed using the standard stereo reconstruction software (MARS)\cite{Ref:MAGICMARS}. This included the calibration, image cleaning and parametrization according to \cite{Ref:MAGICMonoAna} for each MAGIC telescope individually. For moonlight data, a dedicated analysis was performed with optimized image cleaning parameters for the different Night Sky Background (NSB) levels \cite{Ref:MAGICMoon}. Then, a Random Forests (RF) \cite{Ref:MAGICRF} technique with the stereoscopic parameters was used to reconstruct the energy, the arrival direction and to classify the nature of the primary.  Further details about the MAGIC stereo analysis can be found in \cite{Ref:MAGICPerformance}.

In the case of VERITAS, data were analyzed with the one of the standard event reconstruction pipelines, {\tt Eventdisplay}~\cite{Ref:eventdisplay}, and cross-checked with the other standard pipeline, {\tt VEGAS}~\cite{Ref:VEGAS}. In both cases, this included calibration, image cleaning, trace integration, image parameterization, and reconstruction of the energy and arrival direction of gamma-ray-like events.

After separate analysis within the individual collaborations, the data were exported to the DL3 format with the gamma-ray-like events and the associated instrument response functions, and independently analyzed by gammapy \cite{Ref:Gammapy}. Then, the gammapy results were compared with the native ones to check the compatibility between the different analysis pipelines.

\begin{table}[h]
\centering
\resizebox{0.45\textwidth}{!}{%
\begin{tabular}{|c|c|c|}
\hline
\textbf{Year} & \textbf{$\boldsymbol{T_{\mathrm{Eff}}^{\mathrm{MAGIC}}}$ {[}h{]}} & \textbf{$\boldsymbol{T_{\mathrm{Eff}}^{\mathrm{VTS}}}$ {[}h{]}} \\ \hline
2019          & 40                                 & 7                                \\ \hline
2020          & 11                                 & 4                                \\ \hline
2021          & 35                                 & 20                               \\ \hline
2022          & 22                                 & 30                               \\ \hline \hline
2019-2022     & 112                                & 61                               \\ \hline 
\end{tabular}%
}
\caption{Summary of the M87 effective observation time from 2019 to 2022 by MAGIC and VERITAS.}
\label{Tb:Datasets}
\end{table}

\section{Results}
 
 In this section, results from the MAGIC and VERITAS observations corresponding to the period ranging from 2019 to 2022 are reported. This includes the study of the VHE variability by the two instruments and the characterization of the spectral energy distribution (SED) by MAGIC, VERITAS and Fermi-LAT.

\subsection{Light Curves}

\begin{figure}[!h]
\centering\includegraphics[width=1.1\textwidth]{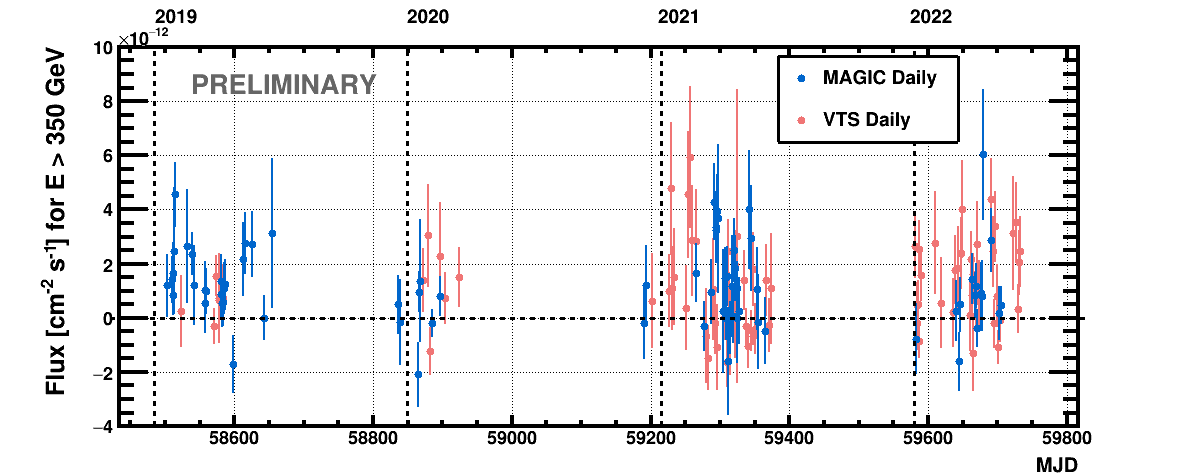}
\caption{Daily light curves from 2019 to 2022 from MAGIC (blue points) and VERITAS (red points).}
\label{Fig:LCDaily}
\end{figure}

\begin{figure}[!h]
\centering\includegraphics[width=1.1\textwidth]{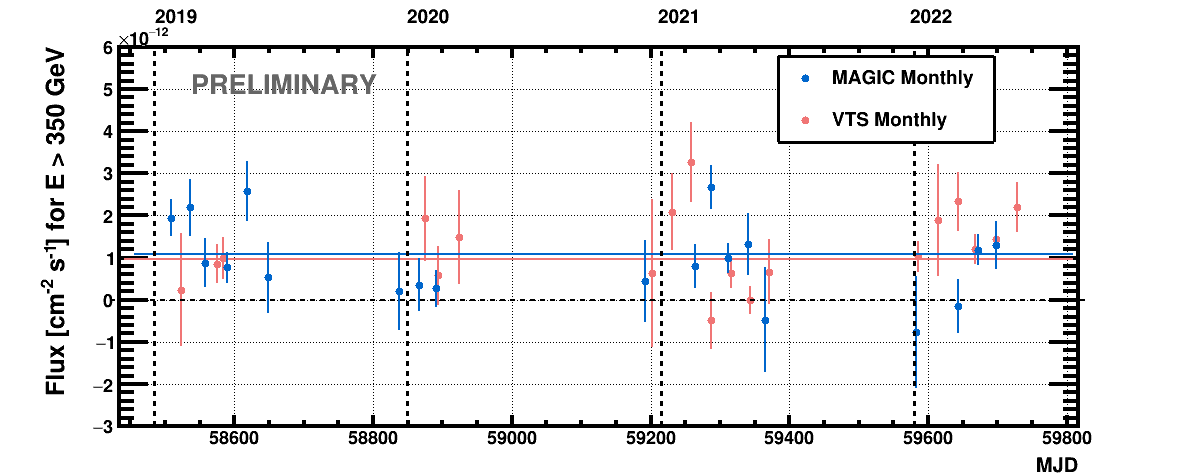}
\caption{Monthly light curves from 2019 to 2022 from MAGIC (blue points) and VERITAS (red points). The mean integral flux is obtained by a fit to a constant to the monthly light curves for MAGIC (blue line) and VERITAS (red line).}
\label{Fig:LCMonthly}
\end{figure}

The variability of the VHE gamma-ray emission of M87 is first studied for energies E > 350 GeV. Results are shown in daily and monthly scales in Figures \ref{Fig:LCDaily} and \ref{Fig:LCMonthly} respectively. The daily and monthly light curves do not show a significant variability for MAGIC and VERITAS observations. This is better investigated by computing the mean integral flux, which is obtained by a fit to a constant to the monthly binned light curves. Results correspond to $(1.07 \pm 0.13)\times 10^{-12}$ $\mathrm{cm^{-2}s^{-1}}$  ($\chi^2/d.o.f=1.98$) and $(0.96 \pm 0.13)\times 10^{-12}$ $\mathrm{cm^{-2}s^{-1}}$ $(\chi^2/d.o.f=1.90)\,$ for MAGIC and VERITAS respectively, as can be seen in Figure \ref{Fig:LCMonthly}. The M87 emission state between 2019 and 2022 is clearly different from the flaring periods in 2005, 2008, and 2010 \cite{Ref:M87MAGIC, Ref:M87VTS, Ref:M87All}, but compatible with previous measurements performed during the low emission state \cite{Ref:M87MAGIC2}.


\subsection{Spectral Energy Distribution}

\begin{figure}[!h]
\centering\includegraphics[width=0.8\textwidth]{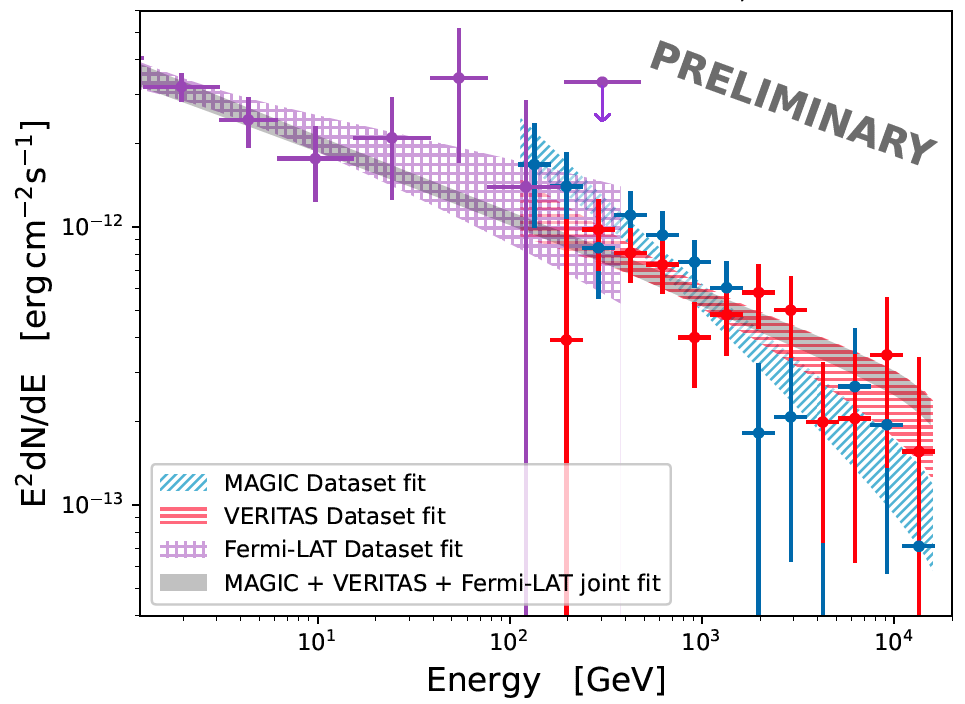}
\caption{Spectral energy distribution between 2019 and 2022 
combining data from MAGIC (blue points), VERITAS (red points) and Fermi-LAT (purple points). The fit to the MAGIC (blue band), VERITAS (red band), and Fermi-LAT (purple band) can be described by a power law. The joint fit (gray band) can be also described by a power-law. }
\label{Fig:SEDJointFit}
\end{figure}

The first joint analysis using combined event data from the two VHE instruments and Fermi-LAT is presented in Figure \ref{Fig:SEDJointFit}. The individual spectra, blue for MAGIC, red for VERITAS, and purple for Fermi-LAT, can be well described by a power-law of the form $E^2dN/dE=f_0(E/E_0)^{\Gamma}$ with $f_0$ the flux normalization, $E_0$ the energy normalization and $\Gamma$ the spectral index. The flux normalization, $f_0$, corresponds to $(3.64 \pm 0.41)\times 10^{-13}$, $(3.66 \pm 0.36)\times 10^{-13}$ and $(1.32 \pm 0.21)\times 10^{-8}$ $\mathrm{cm^{-2}s^{-1} TeV^{-1}}$ for MAGIC, VERITAS and Fermi-LAT respectively.  The energy normalization, $E_0$, is 1 TeV for MAGIC and VERITAS and 10 GeV for Fermi-LAT. Finally, the spectral indices, $\Gamma$, stand for $-2.58 \pm 0.09$, $-2.33 \pm 0.10$ and $-2.19 \pm 0.10$. The observed spectra are not significantly affected by the extragalactic background light (EBL) absorption due to the proximity of M87. 

The joint fit, shown in Figure \ref{Fig:SEDJointFit} as a gray band, spans from below 1 GeV to more than 10 TeV, and can also be described by a power-law with $f_0=(1.25 \pm 0.08)\times 10^{-8}$ $\mathrm{cm^{-2}s^{-1} TeV^{-1}}$, $E_0=10$ $\mathrm{GeV}$ and $\Gamma = -2.28 \pm 0.02$. The spectral index from the joint analysis reported in this work improves the statistical uncertainties obtained with the individual analysis, and it is also in good agreement with previous measurements performed during the low activity state of M87 \cite{Ref:M87MAGIC2,Ref:M87EHT2017}.

\section{Conclusions}

In this work, the M87 observations performed by MAGIC and VERITAS during 2019 to 2022 have been presented. A total of 112 and 61 hours of effective observation was obtained for MAGIC and VERITAS repectively including dark and moonlight conditions. 

First, the light curves for both instruments and for the whole dataset have been shown. No significant variability is observed at daily and monthly scales. Results are clearly different from the three periods of flare activity and compatible with previous low activity state measurements.

Then, the analyses on the individual spectra for MAGIC, VERITAS and Fermi-LAT together with the first joint analysis have been performed with gammapy. Results on the joint analysis show that the data, which spans from below 1 GeV to 10 TeV, can be described by a power-law with index -2.28 $\pm$ 0.02. Results are compatible with previous measurements of M87 during a low activity state and, also, improve the statistical uncertainties obtained with the individual analysis.   


 The extension of the results presented in this work to more years could provide new insights to understand the evolution of the M87 VHE emission, and provide a denser sampling of the source state than what would be obtained for a single experiment.
 
 

\section{Acknowledgments}

We would like to thank the Instituto de Astrof\'{\i}sica de Canarias for the excellent working conditions at the Observatorio del Roque de los Muchachos in La Palma. The financial support of the German BMBF, MPG and HGF; the Italian INFN and INAF; the Swiss National Fund SNF; the grants PID2019-104114RB-C31, PID2019-104114RB-C32, PID2019-104114RB-C33, PID2019-105510GB-C31, PID2019-107847RB-C41, PID2019-107847RB-C42, PID2019-107847RB-C44, PID2019-107988GB-C22 funded by MCIN/AEI/ 10.13039/501100011033; the Indian Department of Atomic Energy; the Japanese ICRR, the University of Tokyo, JSPS, and MEXT; the Bulgarian Ministry of Education and Science, National RI Roadmap Project DO1-400/18.12.2020 and the Academy of Finland grant nr. 320045 is gratefully acknowledged. This work was also been supported by Centros de Excelencia ``Severo Ochoa'' y Unidades ``Mar\'{\i}a de Maeztu'' program of the MCIN/AEI/ 10.13039/501100011033 (SEV-2016-0588, SEV-2017-0709, CEX2019-000920-S, CEX2019-000918-M, MDM-2015-0509-18-2) and by the CERCA institution of the Generalitat de Catalunya; by the Croatian Science Foundation (HrZZ) Project IP-2016-06-9782 and the University of Rijeka Project uniri-prirod-18-48; by the Deutsche Forschungsgemeinschaft (SFB1491 and SFB876); the Polish Ministry Of Education and Science grant No. 2021/WK/08; and by the Brazilian MCTIC, CNPq and FAPERJ.

This research is supported by grants from the U.S. Department of Energy Office of Science, the U.S. National Science Foundation and the Smithsonian Institution, by NSERC in Canada, and by the Helmholtz Association in Germany. This research used resources provided by the Open Science Grid, which is supported by the National Science Foundation and the U.S. Department of Energy's Office of Science, and resources of the National Energy Research Scientific Computing Center (NERSC), a U.S. Department of Energy Office of Science User Facility operated under Contract No. DE-AC02-05CH11231. We acknowledge the excellent work of the technical support staff at the Fred Lawrence Whipple Observatory and at the collaborating institutions in the construction and operation of the instrument.

%
%
%


\newpage
\begin{thebibliography}{99}

\bibitem{Ref:M87Dist}
Mei, S., Blakeslee, J. P., Coté, P., Tonry, J. L., West, M. J., et. al.,
\emph{The ACS Virgo Cluster Survey. XIII. SBF Distance Catalog and the Three-dimensional Structure of the Virgo Cluster},
\href{https://doi.org/10.1086/509598}
{\emph{The astrophysical Journey} \textbf{655} 144 (2007)}

\bibitem{Ref:FRI}
B. L. Fanaroff, J. M. Riley,
\emph{The morphology of extragalactic radio sources of high and low luminosity},
\href{https://doi.org/10.1093/mnras/167.1.31P}
{\emph{MNRAS} \textbf{161} 31-36 (1974)}




\bibitem{Ref:MassM87}
Walter Alef, Keiichi Asada, Rebecca Azulay, Anne-Kathrin Baczko,
\emph{First M87 Event Horizon Telescope Results. I. The Shadow of the Supermassive Black Hole},
\href{https://doi.org/10.3847/2041-8213/ab0ec7}
{\emph{The Astrophysical Journal} \textbf{875} L1 (2019)}

\bibitem{Ref:JetM87Exten}
 Marshall, H. L., Miller, B. P., Davis, D. S., Perlman, E. S., et al.,
\emph{A High-Resolution X-Ray Image of the Jet in M87},
\href{https://doi.org/10.1086/324396}
{\emph{The Astrophysical Journal} \textbf{564} 2 (2002)}

\bibitem{Ref:JetM871}
 Acciari, V. A., Aliu, E., Arlen, T., Bautista, M., et al.,
\emph{Radio Imaging of the Very-High-Energy $\gamma$-Ray Emission Region in the Central Engine of a Radio Galaxy },
\href{https://doi.org/10.1126/science.1175406}
{\emph{Science} \textbf{325} 444 (2009)}

\bibitem{Ref:JetM872}
R. Craig Walker, Philip E. Hardee, Frederick B. Davies, Chun Ly, and William Junor,
\emph{The Structure and Dynamics of the Subparsec Jet in M87 Based on 50 VLBA Observations over 17 Years at 43 GHz},
\href{https://doi.org/10.3847/1538-4357/aaafcc}
{\emph{The Astrophysical Journal} \textbf{855} 128 (2018)}


\bibitem{Ref:JetM87Resol1}
 Eric S. Perlman, John A. Biretta, William B. Sparks, F. Duccio Macchetto, and J. Patrick Leahy ,
\emph{The Optical-Near-Infrared Spectrum of the M87 Jet fromHubble Space Telescope Observations},
\href{https://doi.org/10.1086/320052}
{\emph{The Astrophysical Journal} \textbf{551} 206 (2001)}

\bibitem{Ref:JetM87Resol2}
A. S. Wilson1, and Y. Yang,
\emph{Chandra X-Ray Imaging and Spectroscopy of the M87 Jet and Nucleus},
\href{https://doi.org/10.1086/338887}
{\emph{The Astrophysical Journal} \textbf{568} 133 (2002)}

\bibitem{Ref:M87Hegra}
F. Aharonian, A. Akhperjanian, M. Beilicke, K. Bernlöhr, et al.,
\emph{Is the giant radio galaxy M87 a TeV gamma-ray emitter?},
\href{https://doi.org/10.1051/0004-6361:20030372 }
{\emph{A$\&$A} \textbf{403} 1 (2003)}

\bibitem{Ref:M87HESS}
F. Aharonian, A. G. Akhperjanian, A. R. Bazer-Bachi, M. Beilicke, et al.,
\emph{Is the giant radio galaxy M87 a TeV gamma-ray emitter?},
\href{https://doi.org/10.1126/science.1134408 }
{\emph{Science} \textbf{314} 5804 (2006)}

\bibitem{Ref:M87MAGIC}
J. Albert, E. Aliu, H. Anderhub, L. A. Antonelli, et al.,
\emph{Very High Energy Gamma-Ray Observations of Strong Flaring Activity in M87 in 2008 February},
\href{https://doi.org/10.1086/592348 }
{\emph{The Astrophysical Journal} \textbf{658} L23 (2008)}

\bibitem{Ref:M87VTS}
V. A. Acciari, M. Beilicke, G. Blaylock, S. M. Bradbury, et al.,
\emph{Observation of Gamma-Ray Emission from the Galaxy M87 above 250 GeV with VERITAS},
\href{https://doi.org/10.1086/587458 }
{\emph{The Astrophysical Journal} \textbf{679} 397 (2008)}

\bibitem{Ref:M87Fermi}
 Abdo, A. A., Ackermann, M., Ajello, M., Atwood, W. B., et al.,
\emph{Fermi Large Area Telescope Gamma-Ray Detection of the Radio Galaxy M87},
\href{https://doi.org/10.1088/0004-637X/707/1/55}
{\emph{The Astrophysical Journal} \textbf{707} 1 (2009)}


\bibitem{Ref:M87All}
A. Abramowski, F. Acero, F. Aharonian, A. G. Akhperjanian, et al.,
\emph{THE 2010 VERY HIGH ENERGY $\gamma$-RAY FLARE AND 10 YEARS OF MULTI-WAVELENGTH OBSERVATIONS OF M 87},
\href{https://doi.org/10.1088/0004-637X/746/2/151 }
{\emph{The Astrophysical Journal} \textbf{746} 151 (2010)}

\bibitem{Ref:M87MAGIC2}
 V A Acciari, S Ansoldi, L A Antonelli, A Arbet Engels, C Arcaro, et al.,
\emph{Monitoring of the radio galaxy M 87 during a low-emission state from 2012 to 2015 with MAGIC},
\href{ https://doi.org/10.1093/mnras/staa014 }
{\emph{MNRAS} \textbf{492} 4 (2020)}

\bibitem{Ref:M87EHT2017}
The EHT MWL Science Working Group et al.,
\emph{Broadband Multi-wavelength Properties of M87 during the 2017 Event Horizon Telescope Campaign},
\href{ https://doi.org/10.3847/2041-8213/abef71 }
{\emph{The Astrophysical Journal} \textbf{911} L11 (2021)}

\bibitem{Ref:MAGICPerformance}
Aleksić, J., Ansoldi, S., Antonelli, L. A., et al.,
\emph{The major upgrade of the MAGIC telescopes, Part II: A performance study using observations of the Crab Nebula},
\href{https://doi.org/10.1016/j.astropartphys.2015.02.005}
{\emph{Astroparticle Physics} \textbf{72} 76-94 (2016)}

\bibitem{Ref:VTSPerformance}
Adams, C.B., Benbow, W., Brill, A., Buckley, J. H., et al.,
\emph{The throughput calibration of the VERITAS telescopes},
\href{https://doi.org/10.1051/0004-6361/202142275 }
{\emph{Astronomy $\&$ Astrophysics} \textbf{658} A83 (2022)}

\bibitem{Ref:MAGICMARS}
R. Zanin, et al.,
\emph{MARS, the MAGIC analysis and reconstruction software},
\href{}
{\emph{Proc. of the 33rd ICRC, Rio de Janeiro} \textbf{2} 773 (2013)}

\bibitem{Ref:MAGICMonoAna}
A. M. Hillas,
\emph{Cerenkov light images of EAS produced by primary gamma},
\href{}
{\emph{Proc. of the 19th ICRC, La Jolla} \textbf{3} 445 (1985)}

\bibitem{Ref:MAGICMoon}
M. L. Ahnen, S. Ansoldi, L. A. Antonelli, C. Arcaro, et al.,
\emph{Performance of the MAGIC telescopes under moonlight},
\href{https://doi.org/10.1016/j.astropartphys.2017.08.001}
{\emph{Astroparticle Physics} \textbf{94} 29-41 (2017)}

\bibitem{Ref:MAGICRF}
L. Reiman,
\emph{Random Forest},
\href{https://doi.org/10.1023/A:1010933404324}
{\emph{Machine Learning} \textbf{45} 5-32 (2001)}

\bibitem{Ref:eventdisplay}
Maier, G. and Holder, J.,
\emph{Eventdisplay: An Analysis and Reconstruction Package for Ground-based Gamma-ray Astronomy},
\href{https://ui.adsabs.harvard.edu/abs/2017ICRC...35..747M}
{\emph{Proceedings of the 35th International Cosmic Ray Conference (ICRC 2017)} \textbf{301} 747 (2017)}

\bibitem{Ref:VEGAS}
Cogan, P.,
\emph{VEGAS, the VERITAS Gamma-ray Analysis Suite},
\href{https://ui.adsabs.harvard.edu/abs/2008ICRC....3.1385C}
{\emph{Proceedings of the 30th International Cosmic Ray Conference (ICRC 2007)} \textbf{3} 1385 (2007)}

\bibitem{Ref:Gammapy}
Deil, C., Zanin, R., Lefaucheur, J., et al.,
\emph{Gammapy - A prototype for the CTA science tools},
\href{https://doi.org/10.22323/1.301.0766}
{\emph{Proc. of the 35th ICRC, Busan } \textbf{301} 766 (2017)}


\end{thebibliography}
\end{document}